\documentclass[12pt,a4paper,final]{iopart}

\usepackage{iopams,comment}
\usepackage{iopams}
\usepackage{amsthm}
\usepackage[dvips]{graphicx,epsfig}
\usepackage{pst-plot}
\newtheorem{theorem}{Theorem}[section]

\usepackage[breaklinks=true,colorlinks=true,linkcolor=black,urlcolor=black,citecolor=black]{hyperref}

\begin{document}

\title[Fisher metric from relative entropy group]{Fisher metric from relative entropy group}

\author{Ignacio S. Gomez$^{1,2}$}
\address{$^1$Instituto de F\'{i}sica, Universidade Federal da Bahia,
	Rua Barao de Jeremoabo, 40170-115 Salvador--BA, Brasil}
\address{$^2$IFLP, UNLP, CONICET, Facultad de Ciencias Exactas, Calle 115 y 49, 1900 La Plata, Argentina}
\ead{nachosky@fisica.unlp.edu.ar}

\author{Ernesto P. Borges$^{1}$}
\address{$^1$Instituto de F\'{i}sica, Universidade Federal da Bahia,
	Rua Barao de Jeremoabo, 40170-115 Salvador--BA, Brasil}
\ead{ernesto@ufba.br}

\author[cor1]{M. Portesi$^{2}$}
\address{$^2$IFLP, UNLP, CONICET, Facultad de Ciencias Exactas, Calle 115 y 49, 1900 La Plata, Argentina}
\ead{portesi@fisica.unlp.edu.ar}

\begin{abstract}
In this work we consider the Fisher metric which results from the
Hessian of the relative entropy group, that we called \emph{Fisher metric group}, and we obtain
the corresponding ones to the Boltzmann--Gibbs, Tsallis, Kaniadakis and Abe--Borges--Roditi classes.
We prove that the scalar curvature of the Fisher metric group
results a multiple of the standard Fisher one,
with the factor of proportionality given by the local properties of the entropy
group.
For the Tsallis class,
the softening and strengthening of the scalar curvature is illustrated with the $2D$ correlated model,
from which their associated indexes for the canonical ensemble of a pair of interacting harmonic oscillators,
are obtained.
\end{abstract}

\vspace{2pc}
\noindent{\it Keywords}: entropy group, Fisher metric group, universal classes, statistical models
%\maketitle

\section{Introduction}

For several decades, the geometrical thermostatistics has shown to be a powerful
tool for understanding the thermodynamics of systems in equilibrium \cite{Wei75,Wei76,Rup79,Rup95}.
Among their best achievements are the expression
of phenomenological laws in a rigourous mathematical language, along with a deeper
viewpoint from this geometrical structure.
The geometrical approach for
systems in equilibrium, that began with foundational
works by Gibbs \cite{Gib48}, Hermann \cite{Her73},
Mrugala \cite{Mru78}, and Charatheodory \cite{Cha95},
 led to more concise theoretical one (mainly given by Weinhold
 and Ruppeiner geometries)
where a Riemann metric tensor in the space of thermodynamic parameters
is introduced and a notion of distance between macroscopical states
is provided.
In Ruppeiner geometry, the scalar curvature has proven to be
a useful indicator for characterizing
the effective strength of the interactions present in a system \cite{Rup10}.

In parallel to this progress made in geometrical thermostatistics, empirical and numerical
evidences about the difficulties of the Boltzmann-Gibbs statistical mechanics in describing
several phenomena as systems with long-range interactions
\cite{Kal14}, anomalous diffusion \cite{Tsa96},
structures in plasmas \cite{Guo12}, quantum tunneling and chemical kinetics \cite{Aqu17},
cold atoms \cite{Dou06}, maps with a mixed dynamics \cite{Tir16},
mathematical structures \cite{Niv03,Bor04},
among others, showed some status of anomalies or
failures in the standard theory. Nonextensive
statistics provided a satisfactory answer to these issues
(see, for instance, Ref. \cite{tsallis-book}).

In addition, with the aim of providing a unified theoretical framework for constructing
generalized statistics, the notion of the universal entropy group arose \cite{Tem11}.
In this approach the entropy is not postulated
but only its functional form, which is related to the class of interactions to be considered.
Moreover, groups of entropy can be interpreted
as generalized information measures that
allow one to discriminate between two states represented by
probability density distributions (pdf's).
In particular, when the Kullback-Leibler divergence \cite{Kul59} is considered,
for two neighboring
states, its Hessian provides the Fisher metric. In this way, a Riemannian structure can be
given in the space of the parameterized pdf's, the so-called curved statistical manifolds
\cite{Rao45,Ama00}.

Based on the relative entropy group \cite{Tem11} and by means of the metric provided by its Hessian, in this
work we obtain the Fisher metric group. In this sense, the present contribution can be considered
as a generalization of some previous works of ours \cite{Por06,Por07} and a continuation of the recent
 Tempesta's program \cite{Tem11,Tem15,Tem15bis,Tem16,Tem17} of entropy groups theory in the field of
the information geometry.

The paper is structured as follows. In Section 2 we review
properties of the relative entropy and the associated metric distance, along with
some concepts of the
entropy group theory. Then, in Section 3 from the relative entropy group
we obtain its associated Fisher metric group and we give explicit expressions for some
particular cases.
Section 4 is devoted to discuss the scope
%and new insight
of the formalism presented. We propose a generalization of
the Fisher information and the Cram\'{e}r-Rao bound in the context of the
entropy group theory. We prove that the Riemannian structure derived
by the Hessian of the relative entropy group, is given by a metric tensor that
results a multiple of the standard Fisher one,
through a multiplicative factor given by the local properties of the entropy
group. Here we also show that for the Fisher metric group classes, a
softening and strengthening of the scalar curvature can be
performed by means of an adequate choice of the group parameters.
The results are illustrated for the Tsallis class
with the 2D correlated model,
from which the canonical ensemble of a pair of interacting harmonic oscillators
is also analyzed.
Finally, in Section 5 some conclusions and perspectives are outlined.

\section{Preliminaries}
We first recall the definitions and some properties
of the relative entropy and Fisher metric.
%, along with its thermodynamical limit, the Ruppeiner metric.
Then, we review the notion of relative entropy group.

% 0p
\subsection{Relative entropy}

Consider two pdf's $p(x)$ and $q(x)$ [one has $p,q\geq 0$
$\forall x\in X$ and $\int_{X}p(x)dx=1=\int_{X}q(x)dx=1$
of a continuous variable $x\in X$ with $X$ an abstract set (typically a subset
of $\mathbf{R}^n$)]. The relative entropy or \emph{Kullback-Leibler divergence}
between $p$ and $q$ is
defined as
\begin{equation}\label{relative entropy}
D_{KL}(p\|q)=\int_{X}p(x)\ln\frac{p(x)}{q(x)}dx,
\end{equation}
with the conventions $0.\ln \frac{0}{q}=0$, $p\ln \frac{p}{0}=\infty$.
The content of (\ref{relative entropy}) is as follows. Since $H(p)=-\int_{X}p(x)\ln p(x)dx$
is the Shannon entropy associated with
$p(x)$ and $H(p,q)=-\int_{X}p(x)\ln q(x)dx$ represents
the cross entropy between $p(x)$ and $q(x)$ which measures the average uncertainty when an optimized
$p(x)$ is used instead of a true one $q(x)$, then it follows that $D_{KL}(p\|q)=H(p,q)-H(p)$. This
can be interpreted as $D_{KL}(p\|q)$ measuring the information gain achieved
if $p(x)$ is used instead of $q(x)$.

The relative entropy has several interesting properties:
\begin{itemize}
  \item[$(i)$] it is always nonnegative and vanishes iff $p=q$ a.e.: $D_{KL}(p\|q)\geq0$;
  \item[$(ii)$] it is invariant under parameter transformations: if $y=y(x)$ then
  $D_{KL}(p(x)\|q(x))=D_{KL}(p(y)\|q(y))$;
  \item[$(iii)$] it is additive for noncorrelated variables: if $p(x,y)=p_1(x)p_2(y)$ and
  $q(x,y)=q_1(x)q_2(y)$ then
  $D_{KL}(p(x,y)\|q(x,y))=D_{KL}(p_1(x)\|q_1(x))+D_{KL}(p_2(x)\|q_2(y))$;
  \item[$(iv)$] it is convex: $D_{KL}(\lambda p_1+(1-\lambda)p_2\|\lambda q_1+(1-\lambda)q_2)\leq\lambda
  D_{KL}(p_1\|q_1)+(1-\lambda)D_{KL}(p_2\|q_2)$
   for all $\lambda\in[0,1]$.
\end{itemize}

\subsection{Fisher metric}

The relative entropy allows one to define a metric in the statistical manifold,
thus providing a distinguishability measure between two pdf's.
Consider two pdf's
$p(x)=p(x;\theta)$ and $q(x)=p(x;\theta_ 0)$ defined over a set $X\subseteq\mathbf{R}^n$ with
parameters $\theta,\theta_ 0\in\mathbf{R}^m$, being
$\Delta \theta=\theta-\theta_ 0$ vanishingly small so that
\begin{eqnarray}\label{neighboring pdf}
p(x)=q(x)+\sum_{j=1}^{m}\Delta \theta^j\frac{\partial p}{\partial\theta^j}\big|_{\theta_ 0}\nonumber
\end{eqnarray}
Since $D_{KL}(p\|q)$ has an absolute minimum of value zero when $p=q$ then by expanding
up to
second order around $\theta=\theta_ 0$ one has
\begin{equation}\label{hessian relative entropy}
D_{KL}(p\|q)=\frac{1}{2}\sum_{j,k=1}^{m}\Delta \theta^j\Delta \theta^kg_{jk}(\theta_ 0)+\mathcal{O}(\Delta \theta^3)
\end{equation}
where $g_{jk}(\theta_ 0)$ is a symmetric and positive definite matrix which defines the \emph{Fisher metric}
 (Fisher information matrix) in the space of parameterized pdf's $p(x;\theta)$. Given a
 family of parameterized pdf's
 $p(x;\theta)$, the Fisher metric takes the form
\begin{equation}\label{fisher metric}
g_{jk}(\theta)=\int_{X}\frac{\partial \ln p(x;\theta)}{\partial\theta^j}
\frac{\partial \ln p(x;\theta)}{\partial\theta^k}p(x;\theta)dx
=\left\langle \frac{\partial \ln p(x;\theta)}{\partial\theta^j}\frac{\partial
	\ln p(x;\theta)}{\partial\theta^k} \right\rangle
\end{equation}
for $j,k=1,\ldots, m$ and where $\langle \ldots \rangle$ stands for the mean value
with respect to $p(x;\theta)$.

The formula (\ref{fisher metric}) finds its physical meaning in the context of statistical
models. When modelling an experiment where the results $x\in X$ (called microspace) are distributed
 by $p(x;\theta)$ and $\theta$ are the parameters (macrospace) to be estimated,
the observed information matrix $J_{jk}$ is defined as the log-likehood of $\theta$ given
that occurs $x$, expressed by the formula
$J_{jk}=-\frac{\partial^2 \ln p(x;\theta)}{\partial\theta^j\partial\theta^k}$. Then,
from (\ref{fisher metric}) it follows that
the Fisher metric is precisely the expected value of the observed information matrix.

\subsection{Entropy group theory}

Relative entropies can be expressed by means of the entropy
group theory as generalized information measures. We give a brief review
containing the concepts required in this contribution. We follow the presentation described in Ref. \cite{Tem11}.

The structure of the entropy group theory allows to give a classification
of statistical systems in terms of universality classes.
If $\mathcal{S}$ is an abstract statistical system and $A,B\subseteq\mathcal{S}$ are two independent subsystems,
then the question posed in this formalism concerns
if one can express the joint entropy $S(p_{A\cup B})$ corresponding to the pdf
of the union of the subsystems, $p_{A\cup B}$, as
an analytic function of
the individual ones $S(p_{A})$ and $S(p_{B})$, in the form
\begin{eqnarray}\label{subsystems}
S(p_{A\cup B})=\Phi(S(p_{A}),S(p_{B})) \ \ \ \
\end{eqnarray}
with $\Phi(x,y)$ a polynomial in the variables $(x,y)$. Tempesta \cite{Tem11} answered
satisfactorily this question
by introducing
the \emph{entropy group functional}
\begin{eqnarray}\label{entropy group}
S_{G}(p)=\int_{X}p(x)G(\ln p(x)^{-1})dx\nonumber
\end{eqnarray}
where $G(t)=\ln_{G}(e^t)$ and $\ln_{G}(x)$ is the logarithm group function.
The polynomial $\Phi$ is defined by
the Lazard universal formal group law as
\begin{eqnarray}\label{lazard}
\Phi(x,y)=G(F(x)+F(y))
\end{eqnarray}
where
\begin{eqnarray}\label{polinomial}
G(t)=\sum_{i=0}^{\infty}c_i \frac{t^{i+1}}{i+1}  \ \ \  \textrm{with} \ \ c_0=1 \ , \ c_i\in \mathbf{Q}
\end{eqnarray}
is the so-called formal group exponential and its inverse
\begin{eqnarray}\label{polinomial2}
F= G^{-1}=\sum_{i=0}^{\infty}\gamma_i \frac{s^{i+1}}{i+1}  \ \ \  \textrm{with} \ \ \ \gamma_i\in \mathbf{Q}
\end{eqnarray}
satisfies $G(F(t))=t$ being $\gamma_0=1, \gamma_1=-c_1, \gamma_2=\frac{3}{2}c_1^2-c_2,\ldots$
An advantage of this formalism is that it
allows one to express the additivity property of
the entropy in a simple and general way. For instance,
by choosing $\Phi(x,y)=x+y$ then from
Eq. ~(\ref{subsystems}) one obtains the
additivity of the Boltzmann entropy.
Also, by letting $\Phi(x,y)=x+y+(1-q)xy$ one recovers the
nonadditivity of Tsallis entropies.
Thus, there is a close relation between the
polynomial $\Phi(x,y)$ chosen and the additivity property of the entropy of the system.
In addition, a generalized product $\otimes_G$ can also be defined as
\begin{eqnarray}\label{product law}
\ln_{G}(x \otimes_G y)=\ln_{G} (x) + \ln_{G} (y) \ \ \ \ \forall \ x,y\in\mathbf{R} \ \ \ \ \nonumber
\end{eqnarray}
which extends the property of the logarithm of a product.
One can obtain different generalized statistics by choosing the appropriate
logarithm group function $G(t)$. Here we present
some particular cases for the sake of illustration:
\begin{itemize}
\item[$(I)$] Boltzmann class: this corresponds to the choice $G_{\mathcal{B}}(t)=t$,
  in which case one obtains
the standard additivity and product laws of real numbers along with the Boltzmann-Gibbs entropy.
\item[$(II)$]  Tsallis class: when one chooses $G_{\mathcal{T}}(t)=
  \frac{e^{(1-q)t}-1}{1-q}$
  (with $q$ a real parameter different from one),
 it is straightforward to show that the $q$--sum, the $q$-product \cite{Niv03,Bor04},
 and the Tsallis entropy \cite{Tsa88} are obtained.
\item[$(III)$] Kaniadakis class: by choosing $G_{\mathcal{K}}(t)=
  \frac{\sinh [(1-q)t]}{(1-q)}$
  (with $q$ a real parameter),
the Kaniadakis sum and product, and the Kaniadakis entropy are recovered \cite{Kan02}.
\item[$(IV)$] Abe-Borges-Roditi class: when $G_{\mathcal{ABR}}(t)=
  \frac{e^{at}-e^{bt}}{a-b}$
  with $a \neq b$ real parameters, the
Abe-Borges-Roditi sum and product, along with the Borges-Roditi family of
biparameterized entropies are obtained \cite{Bor98,Abe03}.
\end{itemize}
As an extension of the Kullback-Leibler relative entropy,
the above construction allows one to define the
\emph{relative entropy group}, in the form
\begin{equation}\label{relative group entropy}
D_{G}(p\|q)=\int_{X}p(x)G\left(\ln\left(\frac{q(x)}{p(x)}\right)^{-1}\right)dx
\end{equation}
Some properties of $D_{G}(p\|q)$ have been studied in \cite{Tem11}. For instance if
one assumes that $xG(\ln x)$
is a convex function then the concomitant relative entropy group
is nonnegative and belongs to the class of
Csisz\'{a}r divergence measures.

\section{Universal classes of the Fisher metric group}

As mentioned, the Fisher metric structure
can be derived from the relative entropy.
We propose here to follow a similar argument applied
to the relative entropy group in order to obtain
the associated metric,
that we call Fisher metric group.
Along the same lines that lead to
Eq. (\ref{hessian relative entropy}), we calculate
the matrix elements $g_{jk}(\theta_ 0)_{G}=
\frac{\partial^2 A(\theta)}{\partial\theta^j\partial\theta^k}|_{\theta_ 0}$
where $A(\theta)=
D_{G}(p(x;\theta)||p(x;\theta_0))$,
assuming the pdf depends on a set of parameters
$\{\theta_1,\ldots,\theta_m\}$, with $\Delta \theta=\theta-\theta_0$ vanishingly small,
and $G$ stands
for entropy group. Doing this, we obtain
\begin{eqnarray}\label{derivatives relative group entropy}
\frac{\partial A(\theta)}{\partial\theta^k}=\int_X \frac{\partial p}{\partial\theta^k}
\left[G\left(\ln \frac{p(\theta)}{p(\theta_0)}\right)+
G^{\prime}\left(\ln \frac{p(\theta)}{p(\theta_0)}\right)\right]dx
\nonumber
\end{eqnarray}
from which it follows
\begin{eqnarray}\label{second derivatives relative group entropy}
&\frac{\partial^2 A(\theta)}{\partial\theta^j\partial\theta^k}=
\int_X \frac{\partial^2 p}{\partial\theta^j \partial\theta^k}
\left[G\left(\ln \frac{p(\theta)}{p(\theta_0)}\right)+
G^{\prime}\left(\ln \frac{p(\theta)}{p(\theta_0)}\right)\right]dx+\nonumber\\
&\int_X p(\theta)^{-1}\frac{\partial p}{\partial\theta^j}\frac{\partial p}{\partial\theta^k}
\left[G^{\prime}\left(\ln \frac{p(\theta)}{p(\theta_0)}\right)+
G^{\prime\prime}\left(\ln \frac{p(\theta)}{p(\theta_0)}\right)\right]dx\nonumber
\end{eqnarray}
Specializing this equation when $\theta=\theta_ 0$ then the first term vanishes
due to $\int_X pdx=1$ and the brackets becomes a constant.
So we obtain
\begin{eqnarray}\label{metric group}
g_{jk}(\theta_ 0)_{G}=\left(G^{\prime}(0)+
G^{\prime\prime}(0)\right)\int_X p(\theta_0)
\frac{\partial \ln p(\theta_0)}{\partial\theta^j}\frac{\partial \ln p(\theta_0)}{\partial\theta^k}dx \nonumber
\end{eqnarray}
In turn,
this expression can be recast in the compact form
\begin{eqnarray}\label{fisher metric group}
g_{jk}(\theta)_{G}=\left(G^{\prime}(0)+
G^{\prime\prime}(0)\right)\left\langle
\frac{\partial \ln p(x;\theta)}{\partial\theta^j}\frac{\partial \ln p(x;\theta)}{\partial\theta^k}
\right\rangle
\end{eqnarray}
which constitutes a generalization of the Fisher metric in the language of the entropy group theory,
that we called \emph{Fisher metric group}.
One can see that $g_{jk}(\theta)_{G}$ is simply
the Fisher metric (\ref{fisher metric})
multiplied by the factor
$(G^{\prime}(0)+G^{\prime\prime}(0))$ which in turn depends on the
local properties of $G(t)$ around $t=0$.

We illustrate the Fisher metric group for some relevant cases.
\begin{itemize}
  \item[$(a)$] Boltzmann class:
  In this case the entropic functional corresponds to the Boltzmann-Gibbs entropy so $G_{\mathcal{B}}(t)=t$,
$G_{\mathcal{B}}^{\prime}(t)=1$ and $G_{\mathcal{B}}^{\prime\prime}(t)=0$. Then,
by replacing this in (\ref{fisher metric group})
one obtains the standard Fisher metric (\ref{fisher metric}), as expected.

\item[$(b)$] Tsallis class:
this one corresponds to $G_{\mathcal{T}}^{\prime}(t)=e^{(1-q)t}$
and $G_{\mathcal{T}}^{\prime\prime}(t)=(1-q)e^{(1-q)t}$
so $G_{\mathcal{T}}^{\prime}(0)=1$ and $G_{\mathcal{T}}^{\prime\prime}(0)=(1-q)$.
By replacing this in (\ref{fisher metric group})
one obtains
the Fisher metric of Tsallis class
\begin{eqnarray}\label{Tsallis fisher metric}
g_{jk}(\theta)_{\tau}=
(2-q)\left\langle\frac{\partial \ln p(x;\theta)}{\partial\theta^j}
\frac{\partial \ln p(x;\theta)}{\partial\theta^k}\right\rangle
\end{eqnarray}
By taking the limit $q\rightarrow1$ in (\ref{fisher metric group})
one recovers the standard Fisher metric, as expected.

\item[$(c)$] Kaniadakis class:
for this class we have
$G_{\mathcal{K}}^{\prime}(t)=\cosh [(1-q)t]$ and
$G_{\mathcal{K}}^{\prime\prime}(t)=(1-q)\sinh [(1-q)t]$.
Thus, $G_{\mathcal{K}}^{\prime}(0)=1$
and $G_{\mathcal{K}}^{\prime\prime}(0)=0$.
By replacing this in (\ref{fisher metric group}) again
one recovers the ordinary Fisher metric.

\item[$(d)$] Abe-Borges-Roditi class:
in this class the logarithm group is more general than
in the previous cases, depending on
 two parameters.
We have $G_{\mathcal{ABR}}^{\prime}(t)=\frac{ae^{at}-be^{bt}}{a-b}$ and
$G_{\mathcal{ABR}}^{\prime\prime}(t)=\frac{a^2e^{at}-b^2e^{bt}}{a-b}$. So
$G_{\mathcal{ABR}}^{\prime}(0)=1$ and $G_{\mathcal{ABR}}^{\prime\prime}(0)=a+b$.
By replacing this in (\ref{fisher metric group})
one obtains
\begin{eqnarray}\label{AbeBorgesRoditi fisher metric group}
g_{jk}(\theta)_{\mathcal{ABR}}=
(1+a+b)\left\langle\frac{\partial \ln p(x;\theta)}{\partial\theta^j}
\frac{\partial \ln p(x;\theta)}{\partial\theta^k}\right\rangle
\end{eqnarray}
This class has the particularity of containing all the previous ones.
Boltzmann class corresponds to
$\{a\rightarrow0,b\rightarrow0\}$, while the Tsallis and Kaniadakis ones
are obtained by choosing $\{a=1-q,b=0\}$ and $\{a=1-q,b=-(1-q)\}$.
\end{itemize}

\section{Scope and new insights}

In this Section we
provide some new perspectives of
the Fisher metric group.
We propose a generalized Cram\'{e}r-Rao
inequality (CRI) along
with its associated Fisher--Rao complexity. We
make a characterization of the Riemann structure in the macrospace
by obtaining the scalar curvature
of the universal classes of the Fisher metric group.
We study
some consequences of the Fisher metric group with regard
to the global information-geometric indicators,
and we show an application for
a 2D-correlated Gaussian model. Then, we illustrate the results
with a pair of interacting harmonic oscillators in the canonical ensemble.

\subsection{Generalized Cram\'{e}r-Rao inequality and
Fisher-Rao complexity}
The CRI
is widely used in estimation theory and statistics, since a
distribution
that satisfies equality is considered to
present
the lowest mean square error.
Let $p(x;\theta)$ be a parameterized pdf
with $x,\theta \in \mathbf{R}$. The CRI
establishes that the product of the variance of $x$ times
the Fisher information $I_F(p)$
is greater or equal to
one, i.e.
\begin{eqnarray}\label{CRB}
\langle (x- \langle x \rangle )^2\rangle I_F(p)\geq 1
\ \ \ \ \textrm{with} \ \ \ \ I_F(p)=\left\langle
\left( \frac{\partial \ln p(x;\theta)}{\partial x}\right)^2\right\rangle
\end{eqnarray}
The Fisher information
%is the trace of the Fisher metric (\ref{fisher metric}) when the relationship between
%the microvariable $x$ and the macrovariable $\theta$ is
%linear, i.e. $y=x+\theta$ is a random variable, and
constitutes a measure of the information content of $p$
and it has been shown to be relevant in applications,
both theoretical as experimental ones \cite{Fis22,Fri98}.
In addition, the inequality (\ref{CRB}) suggest to
define the Fisher-Rao complexity as
%(FRC) $C_{FR}(p)$ by means of the formula
\begin{eqnarray}\label{FRC}
C_{FR}(p)=\langle (\langle x \rangle - x)^2\rangle I_F(p)
\end{eqnarray}
so that another statement of the CRI is the following
\begin{eqnarray}\label{CRB2}
C_{FR}(p)\geq1 \nonumber
\end{eqnarray}
The product in
(\ref{FRC}) illustrates how a statistical complexity $C(p)$
is constructed in the standard way:
by means of the product between a disequilibrium function and a disorder
measure normalized to the maximum value,
in such a way that $0\leq C(p)\leq1$.
Furthermore, it can be shown that
the family of pdf's that minimizes the Fisher-Rao complexity
are precisely the Gaussian ones
within
an axiomatization of the statistical complexity for pdf's \cite{Rud16}.

With the aim to generalize the CRI
and the Fisher-Rao complexity
in the context of the Fisher metric group
we define a generalized Fisher information
$I_F^{(G)}(p)=(G^{\prime}(0)+G^{\prime\prime}(0))I_F(p)$, in accordance with
(\ref{fisher metric group}).
Then, from (\ref{CRB}) and assuming
$G^{\prime}(0)+G^{\prime\prime}(0)\geq0$ (this is satisfied for Boltzmann and Kaniadakis classes,
and for Tsallis and Abe-Borges-Roditi only when $q\leq2$ and $a+b\geq-1$ respectively)
it follows straightforwardly a generalized CRI
\begin{eqnarray}\label{generalized-CRB}
\langle (\langle x \rangle - x)^2\rangle I_F^{(G)}(p)\geq
G^{\prime}(0)+G^{\prime\prime}(0)
\end{eqnarray}
from which by defining its associated generalized Fisher-Rao complexity
\begin{eqnarray}\label{generalized-FRC}
C_{FR}^{(G)}(p)=\langle (\langle x \rangle - x)^2\rangle I_F^{(G)}(p)
\end{eqnarray}
which can be stated, equivalently, as
\begin{eqnarray}\label{generalized-CRB2}
C_{FR}^{(G)}(p)\geq G^{\prime}(0)+G^{\prime\prime}(0). \nonumber
\end{eqnarray}

\subsection{Characterization of the universal classes of the Fisher metric group}

The particular form of the Fisher metric group, given by Eq. (\ref{fisher metric group}),
implies some global consequences about the Riemannian
structure in the macrospace $\theta\in\mathbf{R}^m$.
We begin by recalling some expressions of differential geometry and then we
establish a general formula for the
scalar curvature of the Fisher metric group.

Given a metric $g_{ij}(\theta)$, the Christoffel symbols (of the second kind)
are defined as
\begin{eqnarray}\label{christoffel}
\Gamma^i_{kl}=\frac{1}{2}g^{im}
\left(g_{mk,l}+g_{ml,k}-g_{lm,k}\right)  \ \ \ \ \ \  \textrm{for all} \ \ \  i,k,l=1,\ldots,m
\end{eqnarray}
where $g^{im}$ is the inverse matrix of the metric,
$g_{ij,k}$ is a short notation for $\frac{\partial g_{ij}}{\partial \theta^k}$, and
Einstein's convention is used for repeated indices.
In turn, from the Christoffel symbols one can obtain the Ricci tensor $R_{ij}$ and the
scalar curvature $R$, in the form
\begin{eqnarray}\label{ricci-tensor}
R_{ij}=\Gamma^l_{ij,l}-\Gamma^l_{il,j}+\Gamma^m_{ij}
\Gamma^l_{lm}-\Gamma^m_{il}\Gamma^l_{jm} \ \ \ \ \ \  \textrm{for all} \ \ \  i,j=1,\ldots,m
\end{eqnarray}
and
\begin{eqnarray}\label{curvature}
R=g^{ij}R_{ij}
\end{eqnarray}
Let
$(g_{ij})_{G}$, $(\Gamma^i_{kl})_{G}$, $(R_{ij})_{G}$, $R_{G}$ be
the Fisher metric, Christoffel symbols, Ricci tensor and scalar curvature corresponding
a generic $G$ class.
Due to Eq. (\ref{fisher metric group}) one has that $(g_{ij})_{G}$ is proportional to
$(g_{ij})_{\mathcal{B}}$ (where $\mathcal{B}$ refers to the Boltzmann class),
so now the strategy is to obtain $R_G$ in terms of $R_{\mathcal{B}}$.
From Eq. (\ref{fisher metric group}) one obtains
\begin{eqnarray}\label{curvature2}
&(g_{ij})_{G}=(G^{\prime}(0)+G^{\prime\prime}(0))(g_{ij})_{\mathcal{B}} \nonumber\\
&(g^{ij})_{G}=(G^{\prime}(0)+G^{\prime\prime}(0))^{-1}(g^{ij})_{\mathcal{B}} \nonumber
\end{eqnarray}
so it follows that $(\Gamma^i_{kl})_{G}=(\Gamma^i_{kl})_{\mathcal{B}}$.
Therefore, $(R_{ij})_{G}=(R_{ij})_{\mathcal{B}}$ and
$R_{G}=(g^{ij})_G (R_{ij})_{G}=(G^{\prime}(0)+G^{\prime\prime}(0))^{-1}(g^{ij})_{\mathcal{B}} (R_{ij})_{\mathcal{B}}=
(G^{\prime}(0)+G^{\prime\prime}(0))^{-1}R_{\mathcal{B}}$.
Thus, we arrive to our main result of this contribution, stated by the
following theorem.

\begin{theorem}(Geometrical
characterization of the Fisher metric group)
\label{teorema}

\noindent The Riemannian structure derived from
the relative entropy group $D_G(p||q)$ is
given by the metric tensor
$g_{ij}(\theta)_G=\left(G^{\prime}(0)+G^{\prime\prime}(0)\right)g_{ij}(\theta)_{\mathcal{B}}$
where $g_{ij}(\theta)_{\mathcal{B}}$ is the standard Fisher one,
associated to the Boltzmann class.
Consequently, $(\Gamma^i_{kl})_{G}=(\Gamma^i_{kl})_{\mathcal{B}}$,
$(R_{ij})_{G}=(R_{ij})_{\mathcal{B}}$ and
$R_G=\frac{R_{\mathcal{B}}}{G^{\prime}(0)+G^{\prime\prime}(0)}$
with $G^{\prime}(0)+G^{\prime\prime}(0)\neq 0$.
\end{theorem}
Theorem \ref{teorema} expresses that all the relative entropies
of the trace class form $D_G(p||q)$
have the same statistical manifold structure locally
(i.e. except the Fisher metric, the Christoffel symbols,
the Ricci tensor are all of the same type),
as the one given by the Boltzmann class.
While globally, the scalar curvature $R_G$
 is proportional to the
Boltzmann class one through a positive factor
$(G^{\prime}(0)+G^{\prime\prime}(0))^{-1}$.
This result is expected since the metrics $g_{ij}(\theta)_{\mathcal{B}}$ and $g_{ij}(\theta)_{G}$
are not linearly independent but proportional.
Equivalently, by making an expansion around
$\theta=\theta_0$ of $F(\theta)=D_G(p(\theta||p(\theta_0)))$
and neglecting terms of order $\mathcal{O}(\theta^3)$,
one can say that the classes of the relative entropy group are
essentially all the quadratic forms $F(\theta)$ linearly generated by the element
$\frac{1}{2}g_{ij}(\theta)_{\mathcal{B}}\Delta \theta_i \Delta \theta_j$.

\subsection{Softening and strengthening of information-geometric indicators
for a correlated model}

We
analyze some consequences of Theorem \ref{teorema}
regarding the scalar curvature as a global indicator of chaotic dynamics in macrospace.
We accomplish this,
in the case of a nontrivial correlated statistical model of
lowest
dimensionality in macrospace, given by the following family of pdf's:
\begin{eqnarray}\label{2Dcorrelatedmodel}
&p(x,y;\mu,\sigma,r)=\nonumber\\
&\frac{1}{2\pi\Sigma^2\sqrt{1-r^2}}\exp\left(\frac{-1}{2(1-r^2)}\left[\frac{(x-\mu_x)^2}{\sigma^2}
+\frac{y^2\sigma^2}{\Sigma^4}+\frac{2r(x-\mu_x)y}{\Sigma^2}\right]\right)
\end{eqnarray}
called \emph{the $2D$ correlated model} \cite{Caf13}.
Here the macrospace is $\{(\mu_x,\sigma):\mu_x\in\mathbf{R}, \sigma\in\mathbf{R}_{+}\}$
and $r\in [-1,1]$ is the correlation coefficient between the microspace variables $x$ and $y$.
This model can be useful for study many topics:
suppression of classical chaos by quantization \cite{Caf13},
entanglement via scattering processes \cite{Caf12,Ser07}, etc.
Also, the macroscopic constant $\Sigma^2=\langle (x-\langle x\rangle)^2(y-\langle y\rangle)^2\rangle$
is motivated by the Uncertainty Principle by considering $x$ as the position of a particle
and $y$ its conjugated momentum.
From Eqs. (\ref{fisher metric}), (\ref{christoffel}), (\ref{ricci-tensor}) and (\ref{curvature})
one obtains the Fisher metric and its scalar curvature for the Boltzmann as
\begin{eqnarray}\label{2Dcorrelated-boltzmann-metric}
&(g_{ij})_{\mathcal{B}}^{2Dc}=
\frac{1}{(1-r^2)}\left(\begin{array}{cc}
\frac{1}{\sigma^2} &
0 \\
0 & \frac{4}{\sigma^2}
\end{array} \right) \nonumber\\
&R_{\mathcal{B}}^{2Dc}=-\frac{(1-r^2)}{2}  \ \ \ \ \textrm{with} -1\leq r\leq 1
\end{eqnarray}
where the limit
cases $r=0$ and $|r|\rightarrow1$ correspond to the uncorrelated
$R^{2Du}=-\frac{1}{2}$
and the maximally correlated $R\rightarrow0$ ones respectively.
Now if we apply Theorem \ref{teorema} for
obtaining the Fisher metric group for this model
then it follows that
\begin{eqnarray}\label{2Dcorrelated-universal-metric}
&(g_{ij})_{G}^{2Dc}=
(G^{\prime}(0)+G^{\prime\prime}(0))\frac{1}{(1-r^2)}\left(\begin{array}{cc}
\frac{1}{\sigma^2} &
0 \\
0 & \frac{4}{\sigma^2}
\end{array} \right)  \nonumber\\
& R_{G}^{2Dc}=-\frac{(1-r^2)}{2(G^{\prime}(0)+G^{\prime\prime}(0))}
\end{eqnarray}
This equation is the starting point of some characterizations between correlated and
uncorrelated statistical models. For the sake of simplicity we illustrate with the
Tsallis class but similar arguments
can be applied for the other classes.
For the Tsallis class
we have $G^{\prime}(0)=1$ and $G^{\prime\prime}(0)=1-q$, then
%so from (\ref{2Dcorrelated-universal-metric})
\begin{eqnarray}\label{2Dcorrelated-tsallis-metric}
&(g_{ij})_{q}^{2Dc}=
(2-q)\frac{1}{(1-r^2)}\left(\begin{array}{cc}
\frac{1}{\sigma^2} &
0 \\
0 & \frac{4}{\sigma^2}
\end{array} \right)  \nonumber\\
& R_{q}^{2Dc}=-\frac{(1-r^2)}{2(2-q)}
\end{eqnarray}
Interestingly, by choosing $q$ such that $2-q =1-r^2$ one recovers
the $2D$ uncorrelated model. This corresponds to a value of $q$
equal to
\begin{eqnarray}\label{qsoft}
q_{\textrm{soft}}(r)=1+r^2
\end{eqnarray}
where the suffix stands for ``softening", $1\leq q_{\textrm{soft}}(r) \leq 2$ and
the cases $r=0$ and $|r|\rightarrow1^{-}$ correspond to $q_{\textrm{soft}}=1$ and
$q_{\textrm{soft}}\rightarrow2^{-}$.
In addition, by making $r=0$
and $(2-q)=\frac{1}{(1-r^2)}$ in (\ref{2Dcorrelated-tsallis-metric}) one recovers
the $2D$ correlated model with a value of $q$ equal to
\begin{eqnarray}\label{qstrenght}
q_{\textrm{str}}(r)=\frac{1-2r^2}{1-r^2}
\end{eqnarray}
where the suffix stands for ``strengthening", $-\infty\leq q_{\textrm{str}}(r) \leq 1$ and
the limiting cases $r=0$ and $|r|\rightarrow1$ correspond to $q_{\textrm{str}}=1$ and
$q_{\textrm{str}}\rightarrow-\infty$.

Let us clarify the physical meaning
of $q_{\textrm{soft}}$ and $q_{\textrm{str}}$.
We can start
with the $2D$ correlated model
and calculate
the Fisher metric group for the Tsallis class along with
the curvature, thus obtaining
Eq. (\ref{2Dcorrelated-tsallis-metric}).
Then, by choosing $q=q_{\textrm{soft}}$
we recover
the values corresponding to the standard (Boltzmann)
$2D$ uncorrelated model so
this
can be interpreted as a softening of the $2D$ correlated one.
Reciprocally, if we start
with the $2D$ uncorrelated model
and calculate the Fisher metric group and
the curvature for the Tsallis class,
then by letting $q=q_{\textrm{str}}$
the
$2D$ correlated model
is recovered.
By contrast with the first case, this is interpreted as a strengthening of the $2D$ uncorrelated
model since for an adequate choice of
$q$, effective correlations,
characterized by their metric and curvature, seem to appear.
The operation of softening and strengthening
of the scalar curvature is summarized by the limits
\begin{eqnarray}
&\lim_{q\rightarrow q_{soft}}(g_{ij})_{q}^{2Dc}=(g_{ij})_{\mathcal{B}}^{2Du} \nonumber\\
&\lim_{q\rightarrow q_{soft}} R_{q}^{2Dc}=R_{\mathcal{B}}^{2Du}\nonumber
\end{eqnarray}
and
\begin{eqnarray}
&\lim_{q\rightarrow q_{str}}(g_{ij})_{q}^{2Du}=(g_{ij})_{\mathcal{B}}^{2Dc} \nonumber\\
&\lim_{q\rightarrow q_{str}} R_{q}^{2Du}=R_{\mathcal{B}}^{2Dc}\nonumber
\end{eqnarray}
In Fig. 1 it is shown the interplay between
the $2D$ correlated model and the uncorrelated one, along with
the role played by $q=q_{\textrm{soft}}$ and $q=q_{\textrm{str}}$.
\begin{figure}[ht]\label{fig}
  \centerline{\includegraphics[width=14cm]{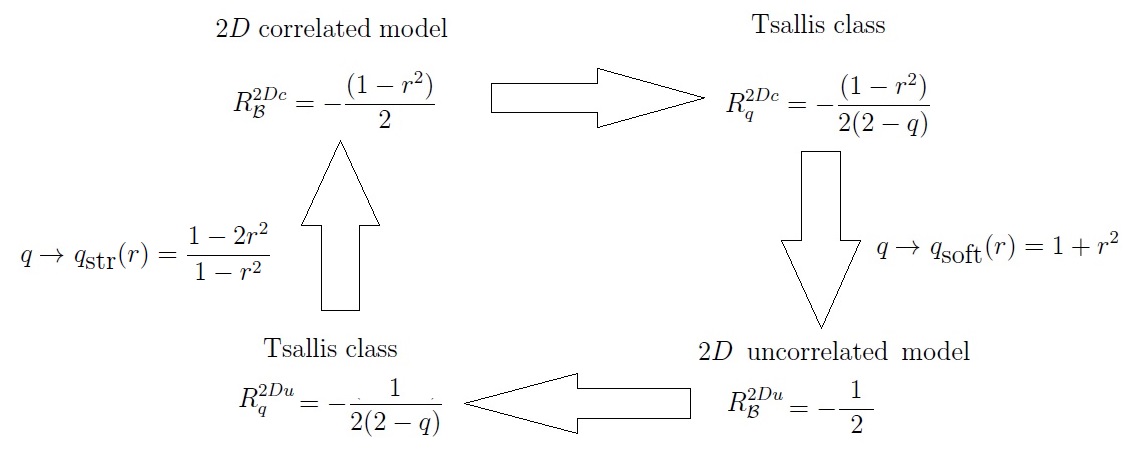}}
  \caption{As a consequence of Theorem \ref{teorema}, it is possible to pass
  from the $2D$ correlated model to the uncorrelated
  one and viceversa, by means of a relationship between
 the correlation coefficient and the index $q$ within the Tsallis class.}
  \end{figure}

\subsection{Application: canonical ensemble description
of two interacting harmonic oscillators}

In order to illustrate the results of this section
we consider a pair of interacting harmonic
oscillators in the canonical ensemble provided
with a total energy $E=T+V$, where
\begin{eqnarray}\label{harmonic-energy}
&T(p_1,p_2)=\frac{p_1^2}{2m_1}+\frac{p_2^2}{2m_2} \nonumber\\
&\textrm{and} \nonumber\\
&V(x_1,x_2)=\frac{1}{2}m_1\omega_1^2(x_1-x_{10})^2+\frac{1}{2}m_2\omega_2^2(x_2-x_{20})^2\nonumber\\
&-r\sqrt{m_1m_2}\omega_1\omega_2(x_1-x_{10})(x_2-x_{20})
\end{eqnarray}
are the kinetic and potential energies respectively.
Here $p_i$, $m_i$, $x_i$, $x_{i0}$, and $\omega_i$
are the momentum, mass, position, equilibrium position,
and frequency of the $i$th-particle
with $i=1,2$. The first two terms of $V$ are the potential energy of
each oscillator separately,
while the third one corresponds to
the interaction term, measuring the coupling strength
in terms of the coefficient $r$.
The canonical ensemble prescription
gives the pdf $p(x_1,x_2,p_1,p_2)$ that characterizes
 the system at equilibrium,
in contact with a thermal bath of temperature $T$, as
\begin{eqnarray}\label{PDF-canonical}
&p(x_1,x_2,p_1,p_2)=A \exp(-\beta E) \nonumber\\
& A^{-1}=\int_{\mathbb{R}^2\times\mathbb{R}^2}
\exp[-\beta E(x_1,x_2,p_1,p_2)]dx_1dx_2dp_1dp_2
\end{eqnarray}
with $\beta=\frac{1}{k_BT}$ the Boltzmann factor. We are only interested in the marginal
PDF $p(x_1,x_2)$ since it contains the 
interaction terms while the momentum marginal 
$\propto \exp[-\beta(\frac{p_1^2}{m_1}+\frac{p_2^2}{m_2})]$ not. Then we can integrate $p(x_1,x_2,p_1,p_2)$ over
$p_1,p_2$, which is equivalent to consider only $V$ instead of $E$ in (\ref{PDF-canonical})
since the kinetic energy does not depend on the coordinates, so we obtain
\begin{eqnarray}\label{marginalPDF-canonical}
&\widetilde{p}(x_1,x_2)=\widetilde{A} \exp(-\beta V) \nonumber\\
&\widetilde{A}^{-1}=\int_{\mathbb{R}^2} \exp[-\beta V(x_1,x_2)]dx_1dx_2 \nonumber
\end{eqnarray}
By means of an adequate choice of the fixed parameters $x_{i0}$, $m_i$, $\omega_i$,
and due to the quadratic form of the potential $V$, the pdf $p(x_1,x_2)$
can be mapped into a $2D$ correlated model (\ref{2Dcorrelatedmodel}). 
Let us define
\begin{eqnarray}\label{constants}
&x_{10}=\mu_x \nonumber\\
&x_{20}=0 \nonumber\\
&\sqrt{k_BT_0(m_1\omega_1)^{-1}}=\sigma\nonumber\\
&\frac{k_B T_0}{\sqrt{m_1m_2}\omega_1\omega_2}=\Sigma \nonumber\\
&1-r^2=\frac{T}{T_0}
\end{eqnarray}
where $T_0$ is a reference temperature for which the interactions are 
absent, typically the room temperature $T_0\sim 293.15 K$. From
Eqs. (\ref{qsoft}), (\ref{qstrenght}) and (\ref{constants})
we obtain the softening and the strengthening indices for the
Tsallis class in terms of the
bath temperature $T$,
\begin{eqnarray}\label{indexes-temperature}
&q_{\textrm{soft}}(T)=2-\frac{T}{T_0} \nonumber\\
&q_{\textrm{str}}(T)=2-\frac{T_0}{T}.
\end{eqnarray}
Hence, $q_\textrm{soft}$ varies linearly with $T/T_0$ while $q_\textrm{str}$ is inversely proportional to $T/T_0$.
From Eq. (\ref{indexes-temperature}) we also can deduce a 
relationship between $q_{\textrm{soft}}(T)$ and $q_{\textrm{str}}(T)$ 
(which is compatible with Eqs. (\ref{qsoft}) and (\ref{qstrenght}))
\begin{eqnarray}\label{indexes-relation}
(2-q_{\textrm{soft}}(T))(2-q_{\textrm{str}}(T))=1
\end{eqnarray}
i.e. they are not independent but linked by the bath temperature.
It is interesting to note how the evolution of the indices as
as the temperature grows from $T=0$ to room temperature $T_0$.
In Table 1 we give $q_{\textrm{soft}}$ and $q_{\textrm{str}}$
for some values of $T/T_0$.

\vspace{0.2truecm}

\begin{table}\label{tabla}
\centering
\begin{tabular}{|l|r|r|}
\hline
%& \multicolumn{2}{c}{Distancia al sol} \\
temperature ratio $[\frac{T}{T_0}]$ & \multicolumn{2}{c|}{indices} \\ \cline{2-3}
& $q_{\textrm{soft}}$ & $q_{\textrm{str}}$ \\ \hline
0 & 2 & $-\infty$ \\
$10^{-5}$ & 1.99999 & -99998 \\
$10^{-4}$ & 1.9999 & -9998 \\
$10^{-3}$ & 1.999 & -998 \\
$10^{-2}$ & 1.99 & -98 \\
$10^{-1}$ & 1.9 & -8 \\
$0.5$ & 1.5 & 0 \\
$0.75$ & 1.25 & 0.6666 \\
$0.9$ & 1.1 & 0.8888 \\
$1$ & 1 & 1 \\ \hline
\end{tabular}
\centering
\caption{$q_{\textrm{sof}}$ and $q_{\textrm{str}}$
for some values of $T/T_0$ as the system evolves
from strong interactions ($T=0$) to the suppression
of these ones ($T=T_0$).}
\end{table}
From Table 1 we see that for a wide range of 
temperatures ($10^{-5}$ to $10^{-1}$) $q_{\textrm{soft}}$ remains with a constant value $\sim 1.99$ and only from $0.5$
to $1$ it changes significatively. By contrast,
$q_{\textrm{str}}$ varies rapidly, from an infinity value to $1$, as the temperature increases.
Physically, this can be interpreted as if it would be easier to attenuate the correlations between
the pair of oscillators
than to add them to an uncorrelated ensemble of those, as the temperature of the bath increases.

\section{Conclusions}

We have presented the Fisher metric that can be obtained
by means of the Hessian of the relative entropy group, that we called
Fisher metric group, and we have obtained explicit expressions
for the Tsallis, Kaniadakis and Abe--Borges--Roditi classes.
We proved Theorem \ref{teorema}, which states
that all the classes of the Fisher metric group
have a metric tensor that is a scalar multiple of the standard Fisher one, with
the factor of proportionality given
in terms of the first two derivatives of the logarithm group $G(t)$
at $t=0$.
The consequences of Theorem \ref{teorema} are in several aspects.

First, by defining a Fisher information motivated by the
Fisher metric group,
a Cram\'{e}r-Rao inequality was extended with the
lower bound given by the first two derivatives of $G(t)$ at $t=0$.

Second, the softening and strengthening of the scalar curvature
was proved for the Tsallis class, as a way of
suppressing correlations in the $2D$ correlated model as well
as of adding interactions in the $2D$ uncorrelated one. Both
operations have indexes associated $q_{\textrm{sof}}$ and $q_{\textrm{str}}$,
which are dependent on the correlation coefficient $r$,
having the same standard value $q=1$ in absent of correlations for $r=0$,
as expected.

Third, and maybe the most significant from the physical viewpoint,
for a pair of interacting harmonic
oscillators in the canonical ensemble a connection between the softening and strengthening indexes of the Tsallis
class and the temperature, was shown.
It was also observed that $q_{\textrm{sof}}$ and $q_{\textrm{str}}$
are related to each other in such a way that the variation of
$q_{\textrm{sof}}$ is much smaller than that the corresponding to
$q_{\textrm{str}}$, as the temperature increases from zero
up to a reference temperature.
In turn, this can be interpreted, at least in the examined example,
as if it would be easier to attenuate the correlations
of a correlated system than to add them to an uncorrelated one.

Our study reveals that the Fisher metric group is a useful tool
for characterizing statistical models in multiple ways: generalization
of Fisher information measures and complexities associated with entropy groups (Section 4.1.),
generation of correlated models from uncorrelated ones and viceversa (Theorem \ref{teorema} and Fig. 1),
and linking
softening and strengthening of global geometric indicators with macroscopical quantities (Table 1).
These results point a way to
generalize in other geometrical directions, as for instance the Ruppeiner and Weinhold geometries.

%%%%%%%%%%%%%%%%%%%%%%%%%%%%%%%%%%%%%%%%%%%%%%%%%%%%%%%%%%%%%%%%%%%%%%%%%%%%%%%%
\section*{Acknowledgments}
The authors acknowledge support received from the National Institute of Science and
	Technology for Complex Systems (INCT-SC), CONICET
	(at Universidad Nacional de La Plata)
	and from CAPES / INCT-SC (at Universidade Federal da Bahia).

\end{document}